# Convolutional network learning of self-consistent electron density via grid-projected atomic fingerprints

Ryong-Gyu Lee and Yong-Hoon Kim*

*School of Electrical Engineering, Korea Advanced Institute of Science and Technology (KAIST), 291 Daehak-ro, Yuseong-gu, Daejeon 34141, Korea.*

E-mail: y.h.kim@kaist.ac.kr

**Abstract**

The self-consistent field (SCF) generation of the three-dimensional (3D) electron density distribution ($\rho$) represents a fundamental aspect of density functional theory (DFT) and related first-principles calculations, and how one can shorten or bypass the SCF loop represents a critical question from both practical and fundamental standpoints. Herein, a machine learning strategy DeepSCF is presented in which the map between the SCF $\rho$ and the initial guess density ($\rho_0$) constructed by the summation of neutral atomic densities is learned using 3D convolutional neural networks (CNNs). High accuracy and transferability of DeepSCF are achieved by expanding the input features to include atomic fingerprints beyond $\rho_0$ and encoding them on a 3D grid. The prediction of the residual density ($\delta\rho$) rather than $\rho$ itself is targeted, and, since $\delta\rho$ corresponds to chemical bonding information, a dataset of small-sized organic molecules featuring diverse bonding characters is adopted. After enhancing the fidelity of the method by subjecting the atomic geometries in the dataset to random strains and rotations, the effectiveness of DeepSCF is finally demonstrated using a complex large carbon nanotube-based DNA sequencer model. This work evidences that the nearsightedness in electronic structures can be optimally represented via the local connectivity in CNNs.

## 1. INTRODUCTION

The mean field or self-consistent field (SCF) method represents a ubiquitous computational approach to deal with complex scientific problems formulated as several coupled differential equations and appear in a wide range of contexts such as the Landau theory for phase transitions[1], Bogoliubov-de Gennes equations for superconductivity [2], Gummel's equations for semiconductor devices[3], and Kohn-Sham density functional theory (DFT) for ab initio electronic structure calculations[4-5], to name a few. In the case of DFT, which has become the standard computational tool for a wide range of science and engineering fields, the SCF solution of the Kohn-Sham (KS) equations identifies the three-dimensional (3D) electron density and at the same time variationally minimizes the total energy. Behind the success of DFT lies the Hohenberg-Kohn theorem, which simplifies the complex many-electron problem to the prediction of the 3D ground-state electron density $\rho(\vec{r})$. Despite its success, the applicability of DFT calculations is typically limited to a few hundred to thousand atoms due to the cubic scaling of the computational cost with respect to the number of atoms. To enable large-scale DFT calculations, advanced algorithms to accelerate SCF cycles have been devised[6] and in parallel several routes to reformulate DFT in the context of order-N and orbital-free DFT methods have been also explored[5, 7-8]. More recently, machine learning (ML) techniques have emerged as promising alternatives, and much efforts are presently being devoted to develop the ML strategies that bypass the iterative solutions of KS equations by predicting the electron density[9-22] or local density of states[23-24] or the KS Hamiltonian[25-31]. The $\rho(\vec{r})$ prediction approach, which arguably represents the most straightforward and flexible route in view of the basis set dependence and other implementation considerations, can be broadly categorized into the models based on basis functions[9, 13, 16, 18-19, 21] and those based on grid representations[10-12, 14-15, 17, 20, 22]. The typically atom-centered basis functions-based models generate $\rho(\vec{r})$ by expanding it in terms of basis functions together with the coefficients learned by fitting the dataset. On the other hand, grid-based models learn the map between input fingerprints and the $\rho(\vec{r})$ represented on a 3D real-space mesh. However, current methods still suffer from several shortcomings: The performance of the former strongly depends on the quality of basis functions and the models developed for a specific set of basis functions would be less transferable to other basis functions. For the latter group, the currently implemented approaches attempted to construct neural networks attached to each grid point, requiring huge computational resources and quickly becoming impractical as the system size increases. Importantly, modern state-of-the-art computer vision approaches such as the convolutional neural network (CNN) have not been applied yet.

In this work, partly motivated by our earlier work on the multigrid acceleration of the SCF cycle within DFT[32-33], we develop the DFT DeepSCF scheme that learns the map between atomic fingerprints encoded on a 3D mesh and the converged SCF density based on CNN model (Figure 1). One central implication of the nearsightedness of electronic structure asserted by Kohn is that local properties like chemical bonding and functional groups are transferable from one environment to another[34], and we hypothesized that this nearsightedness principle could be particularly well mapped to the spatial locality (or local connectivity) in CNNs[35]. This is to our knowledge the first successful demonstration of directly predicting $\rho(\vec{r})$ using the CNN algorithm, where advanced neural network architectures such as

U-Net is available[36]. To achieve this objective, we encode the structural and chemical fingerprints of target materials into 3D real-space grid by augmenting the summation of neutral atomic densities $\rho_0(\vec{r})$, typically used as the initial guess electron density, with additional grid-projected atomic fingerprints. Given that the density difference $\delta\rho$ between $\rho_0$ and $\rho$ corresponds to the chemical bonding information (in practice, the correspondence is typically approximate due to the modification of neutral atomic densities), we devise a new U-Net-based architecture that includes a skip connection to learn the residual $\delta\rho$ and achieve an enhanced prediction accuracy. In addition, to maximize the model transferability or incorporate diverse chemical bonding information, we train our model by adopting a molecular database that includes different chemical bonding configurations and then training neural networks with randomly strained geometries. Since the CNN model does not give equivariant results under the rotation of input features, we additionally enhance the training dataset by randomly rotating molecular geometries. Examining the electronic structures of the test dataset obtained from the single-shot KS diagonalization over the predicted $\rho(\vec{r})$, we confirm that the quality of our DeepSCF or non-SCF DFT calculations is comparable to that of fully SCF DFT counterparts. Finally, we demonstrate the size-extensibility and transferability of the DeepSCF model with butane, crystalline polyethylene, graphene, and composite DNA-carbon nanotube structures.

## 2. RESULTS AND DISCUSSION

### 2.1 Key features of the DeepSCF model

In **Figure 1B**, we schematically show the framework of the DeepSCF model that learns the $\rho_0$ to $\rho$ map based on the CNN. We first project on a 3D mesh the atomic orbital-related fingerprints, which encode the atomic geometry $\mathcal{R}$ and chemical composition $\xi$ and at the same time effectively differentiate various chemical environment. Then, these input features are applied to the CNN model and converted into the final product $\tilde{\rho}_{\text{ML}}$. As the CNN architecture, we employ the U-Net that consists of encoding and decoding blocks involving the repeated sets of the convolution layers and rectified linear unit (ReLU) operation[36]. The encoding (decoding) blocks utilize the max-pooling (deconvolution) operations to decrease (increase) the resolution of hidden features. In addition, to integrate the output features from encoding block into decoding block, these blocks were linked by skip-connections (dotted arrows) (**Figure 1C**; see **Figure S1** and **Table S1** for detail). As will be detailed later, rather than directly predicting $\tilde{\rho}_{\text{ML}}$, we used the U-Net block to generate the residual $\delta\tilde{\rho}_{\text{ML}}$ and added it to $\rho_0$ and normalized $\rho_0 + \delta\tilde{\rho}_{\text{ML}}$ by the total number of electrons to prepare the final feature $\tilde{\rho}_{\text{ML}}$. The electronic structure is then obtained by a single-shot KS diagonalization over the $\tilde{\rho}_{\text{ML}}$ without the SCF loop.

In evaluating the prediction accuracy, since different systems $i$ contain different numbers of electrons $N_e^i$, we introduced the absolute percentage error ($\mathcal{E}_p$)

$$\mathcal{E}_p(\%) = 100 \times \frac{\int d\vec{r}|\rho(\vec{r}) - \tilde{\rho}_{\text{ML}}(\vec{r})|}{\int d\vec{r}\rho(\vec{r})}. \quad (1)$$

In addition, the average prediction accuracy for dataset including several structures was estimated using the weighted formula, i.e. $\langle \mathcal{E}_p \rangle(\%) = 100 \times (\sum_i^N N_e^i \mathcal{E}_p^i)/N_e$, where the $N_e$ is the number of electrons in total dataset.

### 2.2 Input and output features

We next discuss in more detail the input and output features of DeepSCF, which we designed to capture the essential features of the $\rho_0(\vec{r})$ to $\rho(\vec{r})$ map (**Figure 2**). To apply the CNN algorithm for first-principles electronic structure calculations, for theoretical and practical reasons, we propose to project atom-originated information on a 3D real-space grid. Various well-tested atomistic quantities are already available in established DFT codes, and here we employed the SIESTA package based on the numerical atomic orbital basis set[37] and adopted (i) the summation of atomic electron densities $\rho_0(\vec{r}) = \sum_I \rho_I^{atom}(\vec{r})$, where $\rho_I^{atom}$ is neutral atomic densities of separated atom $I$, (ii) the diffuse ion charge density

$$\rho_{ion}(\vec{r}) = -\frac{1}{4\pi}\sum_I \nabla^2 V_I^{local}(\vec{r}), \quad (2)$$

where the $V_I^{local}$ is the local pseudopotential of atom $I$, and (iii) the atomic orbital overlap density

$$\rho_s(\vec{r}) = \sum_{\mu\nu} \phi_\nu^*(\vec{r})\phi_\mu(\vec{r}), \quad (3)$$

where the $\mu$, $\nu$ are the labels for the atomic basis orbitals, projected on a grid with a uniform grid spacing of 0.25 Å. In **Figure 2C**, in predicting the dataset that will be detailed later, we indeed find that, with the increasing number of input fingerprints applied to the initial layers, the error $\langle \epsilon_p \rangle$ of trained models decrease for both training and validation sets. This result shows that each input features contribute to capture the final target, demonstrating their fidelity in encoding the 3D electronic structure.

In terms of the learning target of DeepSCF, we considered in addition to the density $\rho$ itself the learning of the residual density $\delta\rho(\vec{r}) = \rho(\vec{r}) - \rho_0(\vec{r})$ and confirmed several benefits of selecting $\delta\rho$ as the target of ML prediction[38]. Numerically, the $\delta\rho$ is more smoothly distributed on a spatial grid compared to the $\rho$ (compare, e.g., **Figure 3B,C**), effectively reducing the grid spacing and thus providing the computational efficiency to train the model. Moreover, the $\delta\rho$ physically corresponds to the electron transfer between constituent atoms, reflecting the chemical bond information. To validate that the learning of $\delta\rho$ instead of $\rho$ allows more accurate prediction of various chemical environments, we prepared the $\rho$-based architecture ① and $\delta\rho$-based architecture ② of the DeepSCF model (**Figure 1E** left panel), in the latter of which the final feature becomes the summation of $\rho_0$ and $\delta\tilde{\rho}_{\text{ML}}$ (dotted arrow in **Figure 2A**). Testing both models, we confirmed that the $\delta\rho$-learning architecture ② gives lower $\langle \epsilon_p \rangle$ values than the $\rho$-learning architecture ① for both training and test datasets (**Figure 1E** right panel; see **Table S2** for detail).

*2.3 Dataset enhancement: Applying random rotation/strain and increasing the dataset size*

Targeting to predict $\delta\rho$ that approximately corresponds to the chemical bond information, another important ingredient of the DeepSCF model is the preparation of a proper dataset that can cover various chemical bonding cases. For this purpose, we used the TABS dataset which consists of molecules including various chemical species[39] (see Experimental Sections for detail). However, while the CNN is equivariant to the translation of input features (**Figure 2A**), it is inherently not equivariant to the rotation of input features. Indeed, upon considering differently oriented guanine structures, we found that the model derived from the original dataset that consists of the molecular geometries lying on the xy-plane results in much larger $\mathcal{E}_p$ for the molecules rotated from the xy-plane (**Figure 3B**, top panel). Moreover, the original dataset only contains the optimized structures, from which we found that the derived ML models become inefficient in predicting the conditions beyond the equilibrium geometries (**Figure S2C**).

To address these problems, we considered the possibilities of randomly rotating and/or straining the ground-state geometries, and additionally increasing the size of dataset. To systematically demonstrate the dataset enhancement effects, in addition to the original ground-state geometry dataset (Original), we prepared the comparison models that were trained on the rotated (R-only) structures and the simultaneously rotated and -2% ~ +2% strained (R+S) structures. In addition, by increasing the R+S dataset size three times, we prepared the augmented rotated and strained (R+S+A) structures. Then, by rotating a guanine molecule, we obtain for the Original, R-only, R+S, and R+S+A models the average $\mathcal{E}_p$ values of 3.18%, 2.82%, 2.59%, and 1.79%, respectively (**Figure 3B**). It should be also noted that, unlike the model trained on the Original dataset (**Figure 3B**, top panel), the enhanced R-only, R+S, and R+S+A models exhibit uniform $\mathcal{E}_p$ distributions over all rotation directions, representing their capability to equivariantly predict differently orientated molecules. Similarly, by straining a guanine molecule with the strain $\epsilon$ ranging from -1 % to +1 %, we again observe that the Original, R-only, R+S, and R+S+A models show the average $\mathcal{E}_p$ values of 3.37%, 2.71%, 2.42%, and 1.80%, respectively (**Figure 3C**). Consequently, by enhancing the dataset by applying random rotation and strain to the original dataset and then increasing the size of dataset, we achieved the rotation-equivariant and highly accurate prediction properties of DeeSCF. We will below exclusively apply the R+S+A model.

*2.3 Model performance*

Now, we discuss the performance of the DeepSCF model in capturing the electronic structures of test dataset (see Methods for details). We confirmed that the $\tilde{\rho}_{\text{ML}}$ fits well to $\rho$ with $\langle\mathcal{E}_p\rangle$ of 1.67 %, which corresponds to 0.99 of the coefficient of determination ($R^2$) score (**Figure S3**). In addition, we obtained the mean-absolute-errors (MAE) of 0.01 eV/atom, 0.14 eV, 0.11 eV, 0.17 eV/Å, 0.17 eV/Å, and 0.17 eV/Å for the predicted total energy ($E_{\text{tot}}$), lowest unoccupied molecular orbital (LUMO), highest occupied molecular orbital (HOMO) levels, atomic forces of *I* atoms in each molecules along x-, y-, and z-directions ($F_x^I$, $F_y^I$, and $F_z^I$), respectively (**Figure 4B,C,D,F**), confirming the excellent accuracy of the $\tilde{\rho}_{\text{ML}}$-based non-SCF calculations in replacing the full SCF DFT calculations. In **Figure S4A**, we discuss the specific cases of pyridinium chloride ($C_5H_6NCl$) and bromopyruvic acid ($C_3H_3BrO_3$) molecules included in the test dataset.

We now demonstrate the size-extensibility and transferability properties of the DeepSCF model, which are essential in predicting unseen materials beyond the molecular training dataset. Due to the adaptation of the CNN architecture, the DeepSCF model can handle any size of input feature and the size-extensibility should be naturally extended to crystalline systems by applying periodic boundary conditions. In **Figure 5**, we present the calculation results obtained for butane, polyethylene, and graphene, which are not present in the dataset. Note that, while butane has a finite structure like other molecular systems in the training dataset, polyethylene and graphene are crystalline systems that are periodically extended along the one- and two-dimensional directions, respectively (**Figure 5A**). Nonetheless, the accuracies of the DeepSCF-predicted $\tilde{\rho}_{\text{ML}}$ as well as the $\tilde{\rho}_{\text{ML}}$-derived density of states (DOS) obtained for periodic polyethylene and graphene comparable to that for isolated butane (**Figure 5B,C**).

Finally, to demonstrate the capability of handling complex large-scale systems that involve a wide range of chemical and bonding environments, we applied the DeepSCF architecture to the 1644-atom nano-carbon electrode-based DNA sequencer model in which a single-stranded DNA (ssDNA) composed of 30 randomly distributed nucleobases is positioned between two capped metallic (6,6) carbon nanotube (CNT) electrodes (**Figure 6A** and see **Experimental Sections** for details)[40-42]. We obtain the $\tilde{\rho}_{ML}$ prediction error $\mathcal{E}_p$ of 2.10 %, which is comparable to that of 1.67 % for the training dataset (**Figure 6B,C**). The performance could be further improved by including within our dataset molecules that contain phosphorus elements (**Figure 6C**, red arrow) and/or weak dispersion interactions (**Figure 6C**, blue arrow). Performing non-SCF calculation based on this $\tilde{\rho}_{ML}$, we also evaluated the DOS and atomic forces $F_I$. We find that the DOS calculated from $\tilde{\rho}_{ML}$ are comparable to DFT data (**Figure 6D**) and moreover the $\tilde{\rho}_{ML}$-derived atomic forces are also in excellent agreement with the DFT results with the MAE of 1.35, 1.07, and 1.26 eV/Å along the x, y and z directions, respectively (**Figure 6E**). Such high accuracy of DeepSCF in deriving electronic structures from $\rho$ suggests the origins of the outstanding performance of ML-based methods to predict higher-level materials properties that include neural network force fields.

## 3. CONCLUSION

To summarize, we introduced the DeepSCF scheme, which leverages the 3D U-Net CNN model to predict the map between 3D grid-projected atomic fingerprints and the SCF $\rho$. The model first encodes the structural and chemical fingerprints of materials on a 3D mesh via the summation of neutral atomic densities ($\rho_0$) as well as other atomic fingerprints, and then uses a U-Net architecture designed to learn the residual density ($\delta\rho$). We have demonstrated that the residual $\delta\rho$ learning model surpasses the counterpart that learns $\rho$ itself in terms of prediction accuracy and attributed the success to the encoding of the chemical bonding information within $\delta\rho$. Given that $\delta\rho$ represents the chemical bonding information, this model was trained on a database of small-sized organic molecules encompassing a diverse range of chemical bonding configurations. Since the CNN model is not rotationally equivariant, we introduced random rotations to the molecular geometries within the dataset. To further increase the transferability and prediction accuracy of DeepSCF, we additionally applied random strains to the equilibrium geometries and finally expanded the dataset size. After verifying that the quality of the non-SCF DFT calculations is comparable to that of fully SCF DFT counterparts, we demonstrated the size-extensibility and transferability of DeepSCF model with butane, polyethylene, graphene and large-scale nano-carbon-based DNA sequencer model. This development not only provides practical guidelines for applying CNNs and related computer vision machine learing methods to accelerate demanding DFT calculations but also establishes the correspondence between the nearsightedness in electronic structure and the spatial locality in CNNs, thereby providing insights into the mechanistic underpinnings of the success of various ML-based atomistic materials simulation strategies.

## Methods

*Dataset*

To train the DeepSCF model, we first create the ground-state geometry dataset by optimizing atomic geometries of molecules in TABS database[39], which consists of 1641 molecules with 24 different functional categories including C, H, N, O, F, S, Cl, and Br atoms, performing the DFT calculation. We then generate the additional datasets for R-only, R+S, and R+S+A models by randomly rotating and straining atomic geometries of the ground-state dataset, and also increasing the size of dataset. In these datasets, each of molecule structures has cubic unit cell with 20 Å lattice parameters. For examining the model performance, guanine, butane, polyethylene and graphene structures were also constructed. The guanine molecule has cubic unit cell with 20 Å lattice parameters. Both butane and polyethylene inside tetragonal unit cell with lattice parameters a, b = 20 Å and c = 15 Å. The polyethylene is periodically connected along the *c*-axis direction. The graphene structure has tetragonal unit cell with lattice parameters a= 30 Å, b= 32 Å and c = 15 Å which is also periodically repeated along the *a-b* plane. The nano-carbon-based DNA sequencer model was also modeled, in which a ssDNA is positioned between two capped (6,6) CNT electrodes. The ssDNA structure consists of backbone and nucleobases: adenine (A), cytosine (C), guanine (G), thymine (T). We insert randomly generated nucleobases (ATTAGCCGAT), repeated three times, into the ssDNA structure containing 984 atoms with C, H, N, O, P chemical species. Each of capped (6,6) CNT electrodes contain 330 carbon atoms. The nano-carbon-based DNA sequencer model structure has tetragonal unit cell with lattice parameters a = 80 Å, b = 32 Å, and c = 96 Å.

*DFT calculations*

For all above datasets, DFT calculations were carried out using the SIESTA package[37] within the generalized gradient approximation[43]. Troullier-Martins-type norm-conserving pseudopotentials[44] were employed with the numerical atomic orbital basis sets of double-ζ-plus-polarization quality. The Monkhorst-Pack $\vec{k}$-point grid of $1 \times 1 \times 1$, $1 \times 1 \times 15$, and $2 \times 1 \times 2$ was sampled in the Brillouin zone for the molecules, polyethylene and nano-carbon-based DNA sequencer models. The atomic geometries of TABS database which were optimized using B3LYP functional[45-47] were reoptimized within GGA-level until the Hellmann–Feynman ionic forces acting on each atom were below 0.02 eV/Å. The posterior electronic structures were obtained from the non-SCF calculations within SIESTA package.

*Model training*

We implemented the DeepSCF model by modifying the U-Net architecture[36] within Pytorch framework[48] (see **Figure S1** for detail). To avoid the overfitting problem, we randomly divided each dataset for the original, R-only, R+S, and R+S+A model into 80% of training set and 20% of test set. Then, we trained the model to training set using the ADAM optimizer[49] with the MSE loss function. We chose the hyperparameters (kernel size and the number of channels in first layers) which show the smallest test errors for the original model trained over 50 epochs (see **Table S1** for detail). Using these optimized hyperparameters, we trained all models (the original, R-only, R+S, and R+S+A) over 100 epochs to reproduce all results of **Figure 3,4,5,6**.

## ACKNOWLEDGEMENTS

This work was supported by the National Research Foundation of Korea (2022K1A3A1A91094293, 2023R1A2C2003816, RS-2023-00253716) and the BK21 Plus in the Korea Advanced Institute of Science and Technology. Computational resources were provided by KISTI Supercomputing Center (KSC-2023-CRE-0476).

## AUTHOR CONTRIBUTIONS

Y.-H.K. formulated and oversaw the project. R.G.L. developed the computational framework and carried out calculations. Y.-H.K. and R.G.L. analyzed the computational results and co-wrote the manuscript.

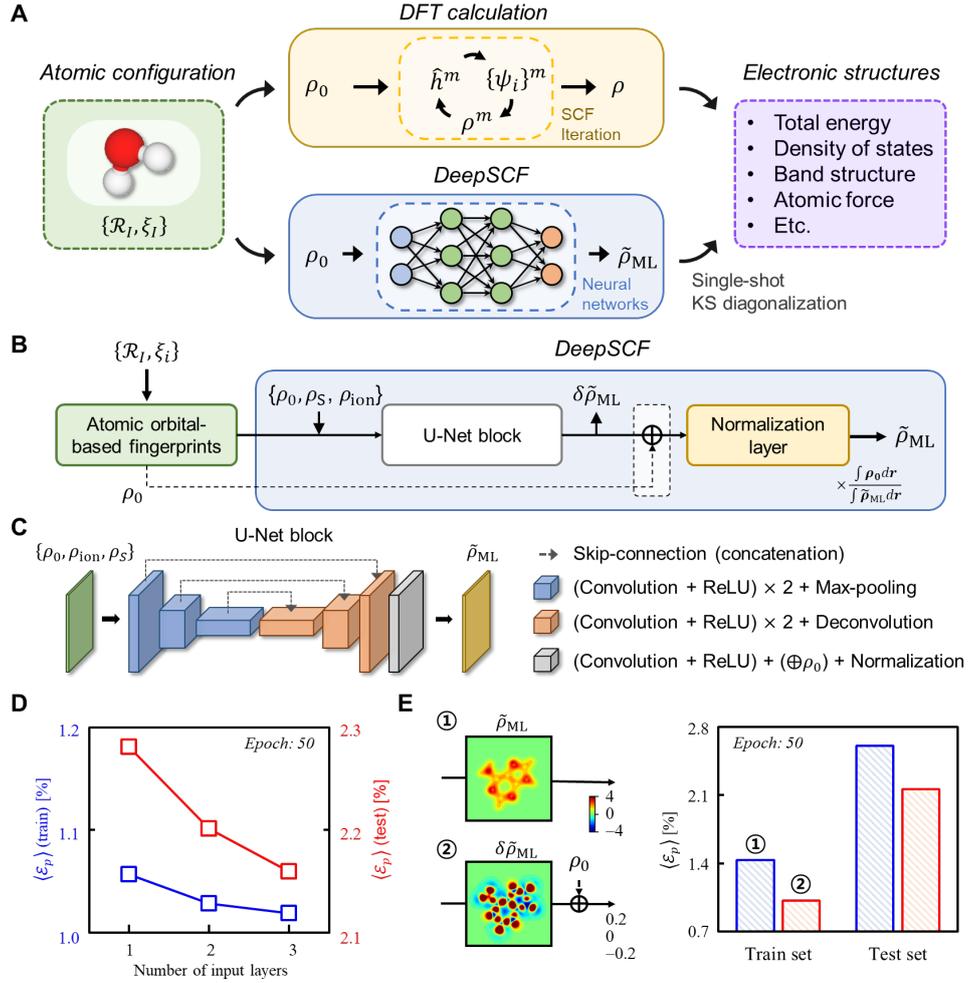

**Figure 1.** Overview of DeepSCF model. A) For material which can be represented by the atomic geometry ($\mathcal{R}_I$) and chemical composition ($\xi_I$) where the $I$ is atomic index, the conventional DFT calculation obtains SCF electron density ($\rho$) through the SCF iterations starting with initial condition of $\rho_0$, in which KS Hamiltonian, wavefunction and electron density are repeatedly calculated until electron density is converged. On the other hands, DeepSCF scheme replaces the SCF iterations with the neural networks to learn the map between $\rho_0$ and prediction target $\tilde{\rho}_{ML}$. For both methods, all ground-state electronic structure properties can be obtained after single diagonalization of the KS equation. B) The atomic orbital-based features extracted from each atomic configuration ($\mathcal{R}$ and $\xi$) are employed into the initial layers of U-Net block of DeepSCF model, and its output $\delta\tilde{\rho}_{ML}$ and $\rho_0$ at initial layer are summed (denoted by $\oplus$) and normalized by $\rho_0$ and thus become final output ($\tilde{\rho}_{ML}$). C) The model's neural networks architecture, including U-Net block which consists of encoder (blue box) and decoder (orange box). Then final (gray box) blocks then receive its output to provide final prediction output $\tilde{\rho}_{ML}$ (see Figure S1 for detail). D) The effects of input atomic orbital-based features on model performance. The 1, 2 and 3 number of input layers respectively represent $\{\rho_0\}$, $\{\rho_0, \rho_{ion}\}$ and $\{\rho_0, \rho_{ion}, \rho_s\}$ features. e, Comparison between $\rho$- and $\delta\rho$-learning approaches which employ different architectures (denoted by ① and ②) at final block in DeepSCF model (dotted box in (A)). The performance of the two approaches are presented by $\langle \mathcal{E}_p \rangle$ for training and test datasets.

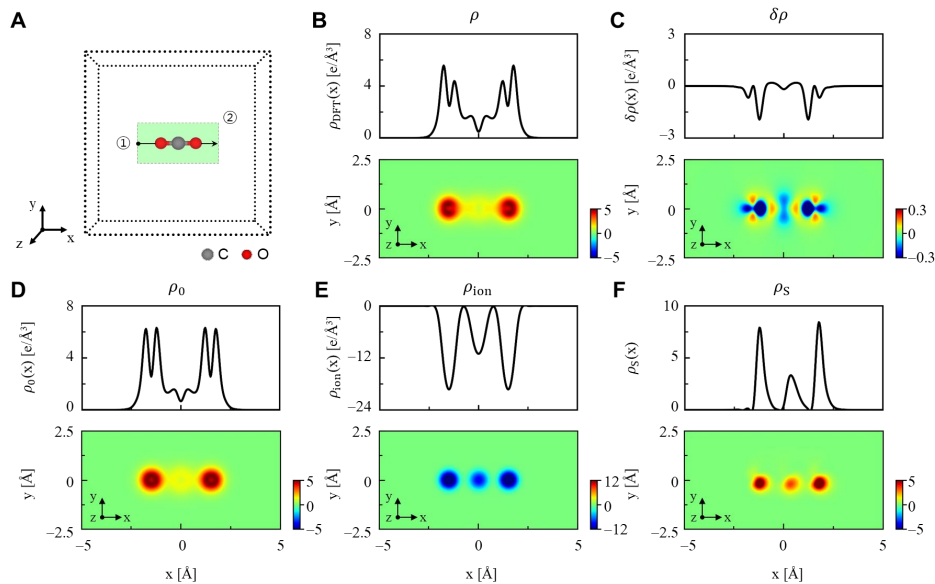

**Figure 2.** Input and output features of carbon dioxide. A) Carbon dioxide model and one- and two-dimensional cross-section (①  and ②) through its center to visualize the spatial distributions of output (B) $\rho$ and (C) $\delta\rho$ and input features (D) $\rho_0$, (E) $\rho_{ion}$ and (F) $\rho_s$.

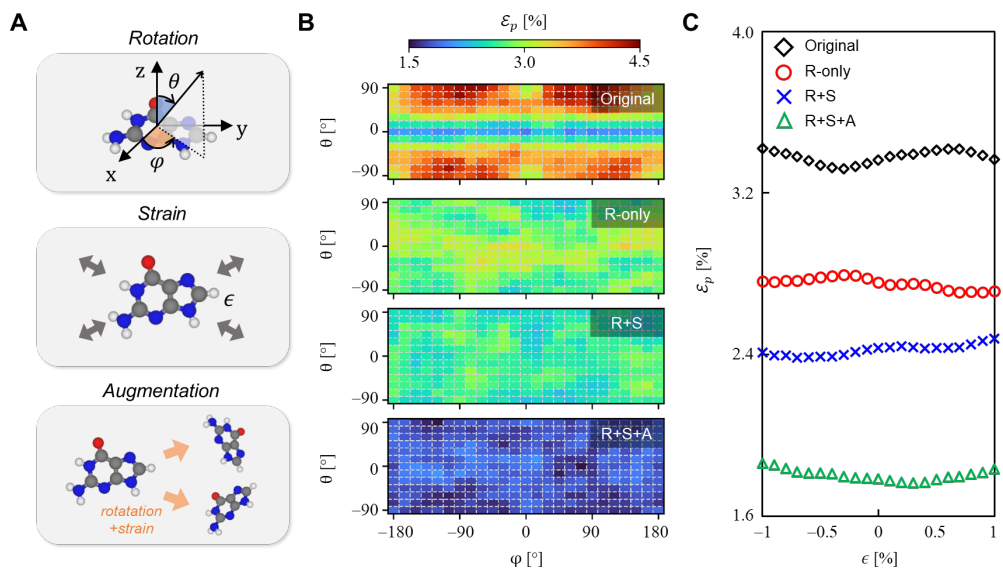

**Figure 3.** Dataset enhancement. A) The Original dataset was enhanced by applying random rotations (θ, φ) in all directions (top) and/or random uniaxial strains (ϵ) ranging from -2% to +2% (middle). The dataset size was further increased by sampling additional rotated/strained geometries (bottom).(B) The $\mathcal{E}_p$ distributions for the rotated guanine structures over the all direction. Top to bottom panels present $\mathcal{E}_p$ distributions of the original, R-only, R+S, and R+S+A models, respectively. C) The $\mathcal{E}_p$ distribution for the guanine structures ($\theta = 0°$, $\varphi = 45°$) strained by the ϵ ranging from -1 to +1%. The black diamond, red circle, blue cross and green triangle indicate the original, R-only, R+S, and R+S+A models, respectively.

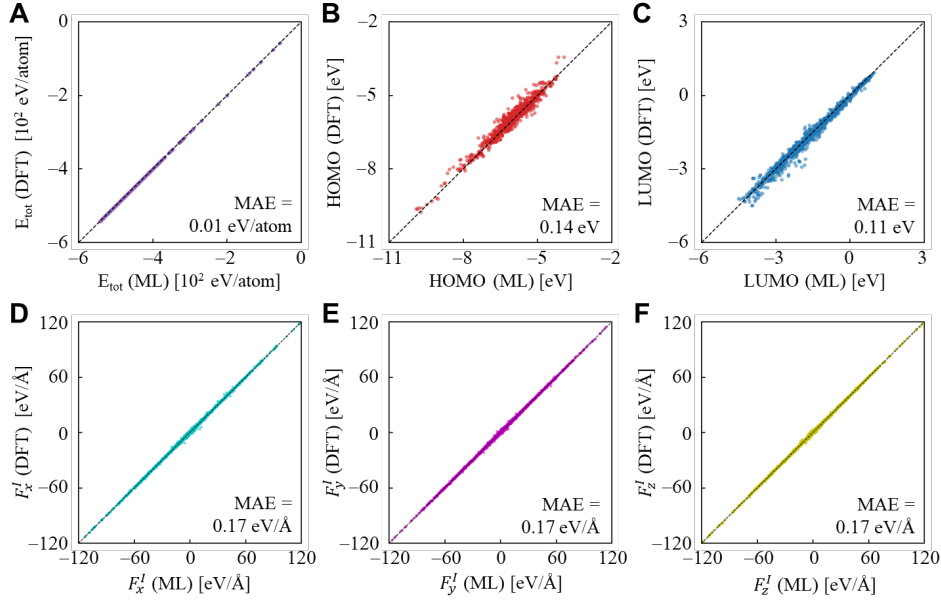

**Figure 4.** Model performance. Performing non-SCF calculations based on the $\rho_{ML}$, the posterior electronic structures of (A) total energy, (B) HOMO, (C) LUMO, (D) $F_x^I$, (E) $F_y^I$, and (F) $F_z^I$ are presented with respect to DFT calculation results producing the MAE of 0.01 eV/atom, 0.14 eV, 0.11 eV, 0.17 eV/Å, 0.17 eV/Å, and 0.17 eV/Å respectively.

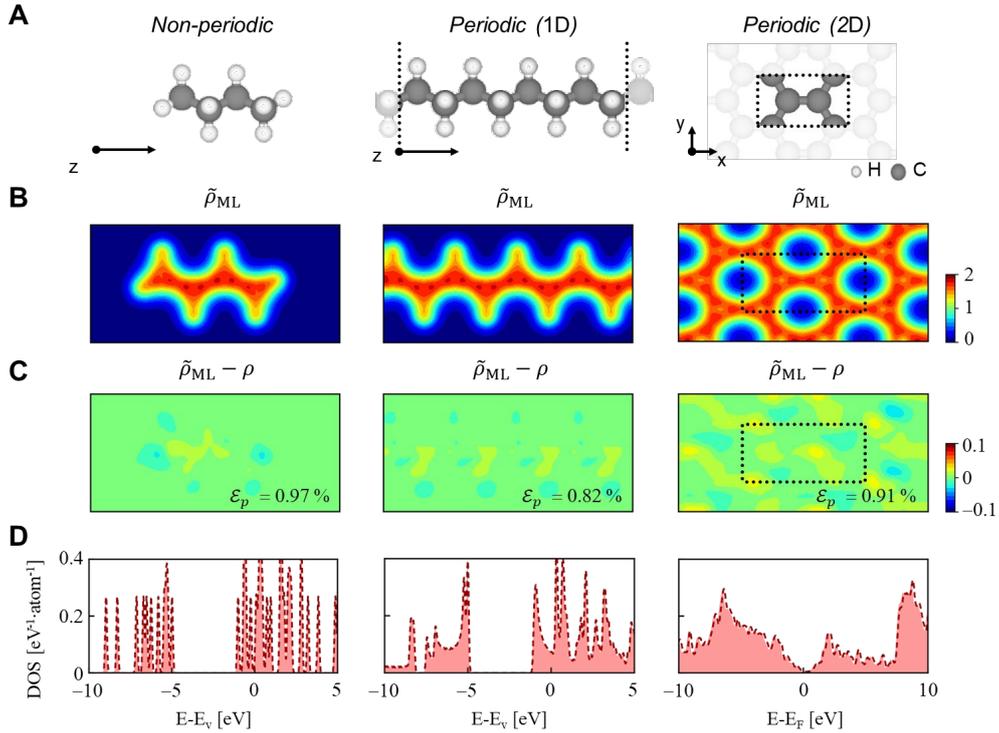

**Figure 5.** Isolated and periodic structures. A) The butane (left), polyethylene (center) and graphene (right) structures. Two dimensional cross-sections of (B) $\rho_{ML}$ and (C) $\rho_{ML} - \rho$ through the center of (A) in x-y and z-x planes, in which color scale is in units of e/Å$^3$. D) The comparison between DOS obtained from $\rho_{ML}$ (red dashed line) and DOS obtained from DFT calculation (red shade) for the all cases.

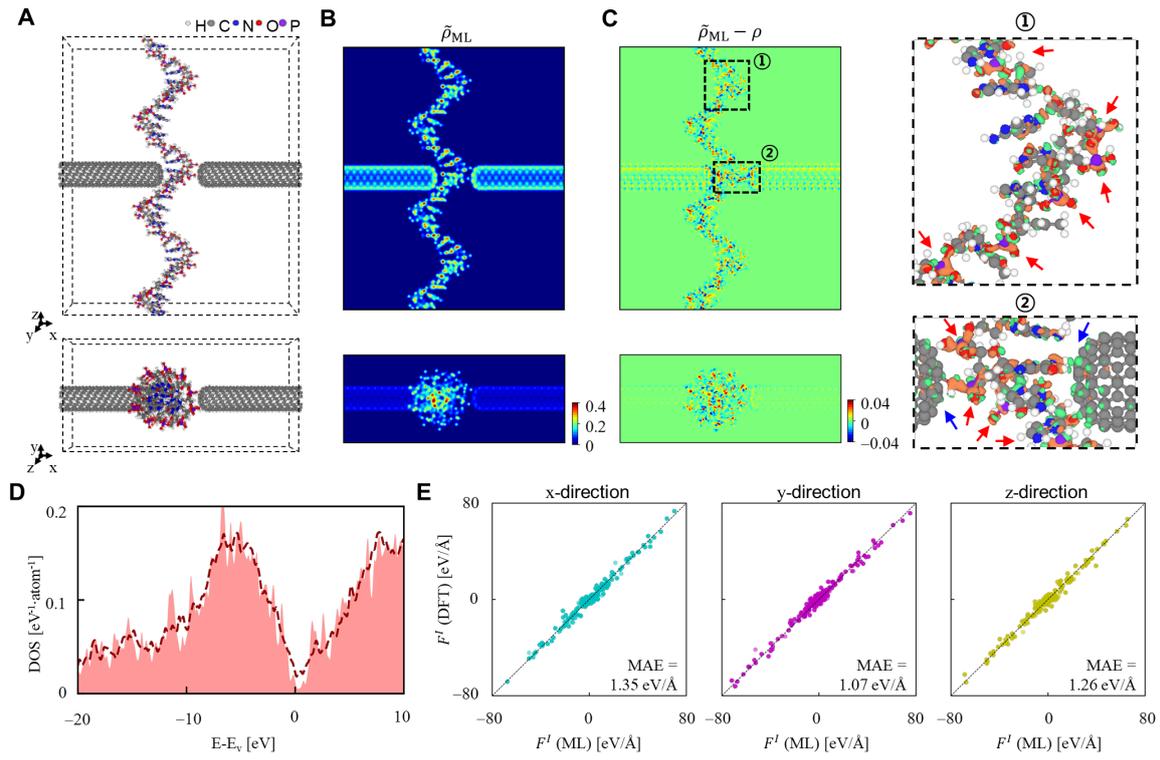

**Figure 6.** The nano-carbon-based DNA sequencer model. A) The ssDNA structure periodically extended along x and z-axis direction which is composed of 30 base pair. B) The plane-averaged $\tilde{\rho}_{ML}$, where color scale is in units of e/Å$^3$. C) The plane-averaged $\tilde{\rho}_{ML} - \rho$ (left panel) and 3D contour plots for $\tilde{\rho}_{ML} - \rho$ in ① and ② regions (right panel), overlaid over the atomic structure. In the ① and ②, red and blue arrows highlight the positions of phosphorous elements and the contacts between the ssDNA and CNT electrode, respectively. The isosurface level for 3D contour plots is 0.03 e/Å$^3$, where the red and green colors indicate positive and negative values. E) The comparison between DOS obtained from $\rho_{ML}$ (red dashed line) and the DFT results (red shade). The comparison between atomic forces of each atoms ($F_I$) obtained from $\tilde{\rho}_{ML}$ and the DFT results along x-, y- and z- directions, respectively.